%% file: cosmosis.tex
\documentclass[5p]{elsarticle}
\pdfoutput=1
\usepackage{graphicx}
\usepackage{xspace}
\usepackage{color}
\usepackage{hyperref}
\usepackage{listings}

\newcommand{\cosmosis}{{\sc CosmoSIS}\xspace} 
\journal{Astronomy \& Computing}

\begin{document}
\begin{frontmatter}

\title{CosmoSIS: modular cosmological parameter estimation}
\author[man]{Joe Zuntz}
\ead{joseph.zuntz@manchester.ac.uk}
\author[fnal]{Marc Paterno}
\author[kavli,enrico]{Elise Jennings}
\author[kavli,rcc]{Douglas Rudd}
\author[kavli,astro]{Alessandro Manzotti}
\author[fermi,kavli,astro]{Scott Dodelson}
\author[man]{Sarah Bridle}
\author[fnal]{Saba Sehrish}
\author[fnal]{James Kowalkowski}

\address[man]{Jodrell Bank Centre for Astrophysics, University of Manchester, Manchester M13 9PL, U.K.}
\address[fnal]{Fermi National Accelerator Laboratory, Batavia, IL 60510-0500, U.S.A.}
\address[kavli]{Kavli Institute for Cosmological Physics, University of Chicago, Chicago, IL 60637, U.S.A}
\address[enrico]{Enrico Fermi Institute, University of Chicago, Chicago, IL 60637, U.S.A}
\address[rcc]{Research Computing Center, University of Chicago, Chicago, IL 60637, U.S.A.}
\address[astro]{Department of Astronomy \& Astrophysics, University of Chicago, Chicago, IL 60637, U.S.A.}

\begin{abstract}
Cosmological parameter estimation is entering a new era. Large collaborations need to coordinate high-stakes analyses using multiple methods; furthermore such analyses have grown in complexity due to sophisticated models of cosmology and systematic uncertainties. In this paper we argue that \emph{modularity} is the key to addressing these challenges: calculations should be broken up into interchangeable modular units with inputs and outputs clearly defined. We present a new framework for cosmological parameter estimation, \cosmosis, designed to connect together, share, and advance development of inference tools across the community. We describe the modules already available in \cosmosis, including {\sc camb}, {\sc Planck}, cosmic shear calculations, and a suite of samplers. We illustrate it using demonstration code that you can run out-of-the-box with the installer available at \url{http://bitbucket.org/joezuntz/cosmosis}.
\end{abstract}

\end{frontmatter}

\input{1-intro}
\input{2-challenges}
\input{3-architecture}

\input{4-demos}

\input{5-sharing}

\bibliographystyle{elsarticle-num}
\bibliography{cosmosis}

\appendix
\input{a1-user-guide}

\input{a2-dev-guide}

\input{a3-example-code}
\input{a4-architecture-details}

\input{a5-technical}
\input{a6-kde}
\input{a7-consistency}

\end{document}

%% file: 1-intro.tex
\section{Introduction}
\label{introduction section}

Cosmological parameter estimation (CPE) is the last step in the analysis of most cosmological data sets. 
After completing all the acquisition, verification, and reduction of the data from an experiment, we transform compressed data sets, such as power spectra or brightnesses of supernovae, into constraints on cosmological model parameters by comparing them to theoretical predictions.

The standard practice in cosmology is to take a Bayesian approach to CPE.  A likelihood function is used to assess the probability of the data that were actually observed given a proposed theory and values of that theory's parameters.  Those parameters are varied within a chosen prior in a sampling process such as Markov Chain Monte-Carlo \cite{2001CQGra..18.2677C}.

The result is a distribution that describes the posterior probability of the theory's parameters, often summarised by the best fit values
and uncertainties on the parameters given the model and data.

A golden age of CPE has just ended; over the past decade the most powerful cosmological probes measured either the background expansion of the universe (like supernovae and baryon acoustic oscillations) or linear perturbations to it (like the microwave background or large-scale structure).  These observations established a powerful concordance model of cosmology, $\Lambda$CDM, in which pressureless dark matter forms the seed and backbone of physical structures and a cosmological constant dominates expansion.

 In the background and linear regimes inter-probe correlations were negligible, statistical errors dominated, and predictions were easy to make.  The challenge for the next decade, for stage III cosmological surveys \cite{2006astro.ph..9591A} and beyond, is to test the $\Lambda$CDM model with data from new regimes.  This will be extremely difficult: on the non-linear scales we must probe, systematic effects require complex modelling, adding new physics becomes much harder, and different probes have subtle statistical correlations.  In this paper we argue that this new era requires a new generation of parameter estimation tools.

A range of CPE tools exists today.  The first and most widely used has been {\sc cosmomc} \cite{2002PhRvD..66j3511L}, a sophisticated Metropolis-Hastings sampler coupled closely to the {\sc camb} \cite{2000ApJ...538..473L} Boltzmann integrator.  Other Boltzmann codes have had their own samplers attached, such as {\sc analyzethis} \cite{2004JCAP...09..003D} for the {\sc cmbeasy} \cite{2005JCAP...10..011D} code and {\sc montepython} \cite{2013ascl.soft07002A} for {\sc class} \cite{2011arXiv1104.2932L}.  Additions to {\sc cosmomc} to perform new kinds of sampling have been made by codes like {\sc cosmonest} \cite{2011ascl.soft10019P}, and methods to interpolate or approximate likelihood spaces have included {\sc cmbfit} \cite{2004PhRvD..69f3005S}, {\sc cmbwarp} \cite{2004PhRvD..70b3005J}, {\sc pico} \cite{2007ApJ...654....2F}, {\sc dash} \cite{2002ApJ...578..665K}, {\sc bambi} \cite{2012MNRAS.421..169G} and {\sc scope} \cite{2014arXiv1403.1271D}.  Other methods, like {\sc fisher4cast} \cite{2011IJMPD..20.2559B} and {\sc icosmo} \cite{2011A&A...528A..33R} have focused on the Fisher matrix approximation, particularly for forecasting results.  More recently, {\sc cosmohammer} \cite{2013A&C.....2...27A} has introduced cloud computing and a more pipeline-like approach to the genre, while {\sc cosmopmc} \cite{2011arXiv1101.0950K} has focused on late-time data with a new sampling algorithm, and {\sc cosmolike} \cite{2014MNRAS.440.1379E} has made great strides in the high-accuracy calculation of the covariances and other statistics of late-time quantities. Codes like {\sc cosmopp} \cite{2013arXiv1312.4961A} and {\sc cosmoslik} have moved towards an object-oriented or plug-in approach to building pipelines.

In this paper we present \cosmosis, a new parameter estimation code 
with \emph{modularity} at its heart, and discuss 
how this focus can help overcome the most pressing challenges 
facing the current and future generations of precision cosmology. 
The goal of \cosmosis is \emph{not} to produce new calculations of physical quantities, but to provide a better way to bring together and connect existing code.  It does this with a plug-in architecture to connect multi-language \emph{modules} performing distinct calculations, and provides a simple mechanism for 
building and extending 
physics, likelihood, and sampler libraries. \cosmosis differs from previous parameter estimation codes in simultaneously emphasizing this modular approach, and allowing cosmologists to develop in their language of choice and thus leverage the large amount of powerful existing code in the community.

In Section~\ref{challenges section} we discuss the challenges brought on by the new generation of data. We describe how modularity addresses many of these challenges in~\ref{modularity section}. We outline the structure of \cosmosis in~\ref{architecture section} and illustrate the \cosmosis features with a number of examples in~\ref{examples section}. We propose a model for future collaborative development of \cosmosis in Section~\ref{sharing section} and wrap up with a discussion of future prospects in~\ref{discussion section}.  Guides for developers and users, and a worked example are included among the appendices.

%% file: 2-challenges.tex
\section{Challenges}

\label{challenges section}

Several problems have conspired to end the pleasant period of CPE.  Cosmological data sets now probe 
a non-linear, multi-probe regime where complex physical and analysis systematics are dominant.  These systematics (such as photometric redshift errors or baryonic effects on the power spectrum) are correlated between probes: we must take care to consistently model their impact on different measurements and inferred statistics.

A richer model accounting for more physics will also require a large increase in the number of parameters.  The Planck mission, for example, required about 20 nuisance parameters to account for physical and instrumental systematics \cite{2013arXiv1303.5076P}. This expanded parameter space carries with it computational costs, and the number of parameters and the associated costs will increase with future experiments.

Since systematics are both dominant and poorly understood, each analysis must be run 
with different models for each systematic to ensure that conclusions are 
robust and insensitive to model choices.
Galaxy bias, for example, which describes the relative clustering of 
a sample of galaxies compared to the underlying matter distribution,
 can be described by a range of different models and parameterizations that are accurate to varying degrees over a given range of scales or galaxy types. 
A computational framework that does 
not allow these models to be simply replaced with alternatives
can quickly become overwhelmingly complicated.

It is not only models of systematics that are getting more complicated. With the rich data expected from current and next generation experiments, 
we will be able to test a wide range of alternatives to vanilla $\Lambda$CDM, such as 
theories of modified gravity or dynamical dark energy (see \cite{2014arXiv1407.0059J} for a recent review).  Many analyses and calibration methods assume $\Lambda$CDM throughout and can make switching to another cosmological model
 very difficult.  Clarifying and making assumptions explicit is vital to correct work in these areas. Moreover, alternative models or parameterizations vary from analyst to analyst, and the most generic of them contains 
dozens of new cosmological parameters 
(for example, the effective Newton's constant as a function of scale and time that relates density to potential), all of which can be constrained.

Since all these complexities can make for a rather slow likelihood evaluation, more advanced sampling methods than the basic Metropolis-Hastings sampling are often considered.  Making it as easy as possible to change and explore sampling methods is therefore a key goal.  Of particular interest are those samplers designed to perform their calculations in parallel which can be used on modern multi-core and multi-node computing platforms.

Many of these problems have been tackled (in code) in a heterogeneous way by multiple authors, with multiple programming languages and in different ways.  A useful CPE framework must make use of the large amount of existing code that was created to tackle different parts of the problems already discussed.

A final problem is social.  Most cosmology collaborations are large and widely geographically spread, making cleanly sharing and comparing code and methods a significant challenge.  There may be multiple approaches to treatment of systematics, multiple ideas for theoretical models to be tested, and multiple preferred computer languages. An easy way to communicate about code development maximizes collaboration between experts at different institutions.

We therefore have a slate of problems: correlated systematics, dimensionality, systematic models, variant cosmologies, advanced sampling, legacy code, 
bug finding, and diverse approaches; these inform our requirements for \cosmosis.

\section{Modularity}

\label{modularity section}

Modularity is the key to solving most of the problems listed above. 

A modular (or \emph{loosely coupled}) approach breaks up a larger complicated code into smaller parts.  The philosophy is then that each module has a specific task to complete - it does one thing, and does it well, and its functionality does not depend directly on what other modules are used in the pipeline - provided that the required inputs are present a module does not care which other code they came from.

The modules are only connected in a specific and limited way - the inputs they take and the outputs they make are passed on only through a specific set of functions designed for this purpose, rather than, for example, creating new global variables or structures to pass around.  They do not have direct read and write access to the data each other hold.\footnote{A modular design is the norm in many areas of software engineering; your web browser and operating system almost certainly take this approach.}.   

All data is then passed around
via a single mechanism - the loading and saving of information in a single place (in this case the datablock; see Section \ref{architecture section}).

A likelihood function then becomes
a sequence of modular processes, run one-by-one to 
form a pipeline.  The last module(s) generates the final likelihood numbers.  Any module in the sequence can be replaced at runtime by another calculation of the same step without affecting the others.
This independence and easy replacement of modules creates a flexible CPE framework where systematic models and alternative cosmologies can be 
fully explored. 

As an example, consider a likelihood calculation for a spectroscopic survey's galaxy power spectrum $P(k,z)$ (see, for example, \cite{2012PhRvD..86j3518P} \cite{2010MNRAS.404...60R}).  We can split the physical calculation into:
\begin{itemize}
	\item Compute the linear matter power spectrum $P(k,z)$ from cosmological parameters using a Boltzmann code.
	\item From this, compute the non-linear $P_{\mathrm{NL}}(k,z)$ with Halofit~\cite{Smith:2002dz} or another model.
	\item Use bias parameters to calculate a bias model $b(k,z)$ for the galaxy sample.
	\item Compute the galaxy power spectrum as  $P_g(k,z) = b^2(k,z)P_{\mathrm{NL}}(k,z)$.
	\item Integrate over survey window functions to compute predictions to compare with measurements.
	\item Compare the predicted to observed measurements to give a likelihood.
\end{itemize}

\begin{figure}[h!]
	\centering
       \includegraphics[width=0.5\textwidth]{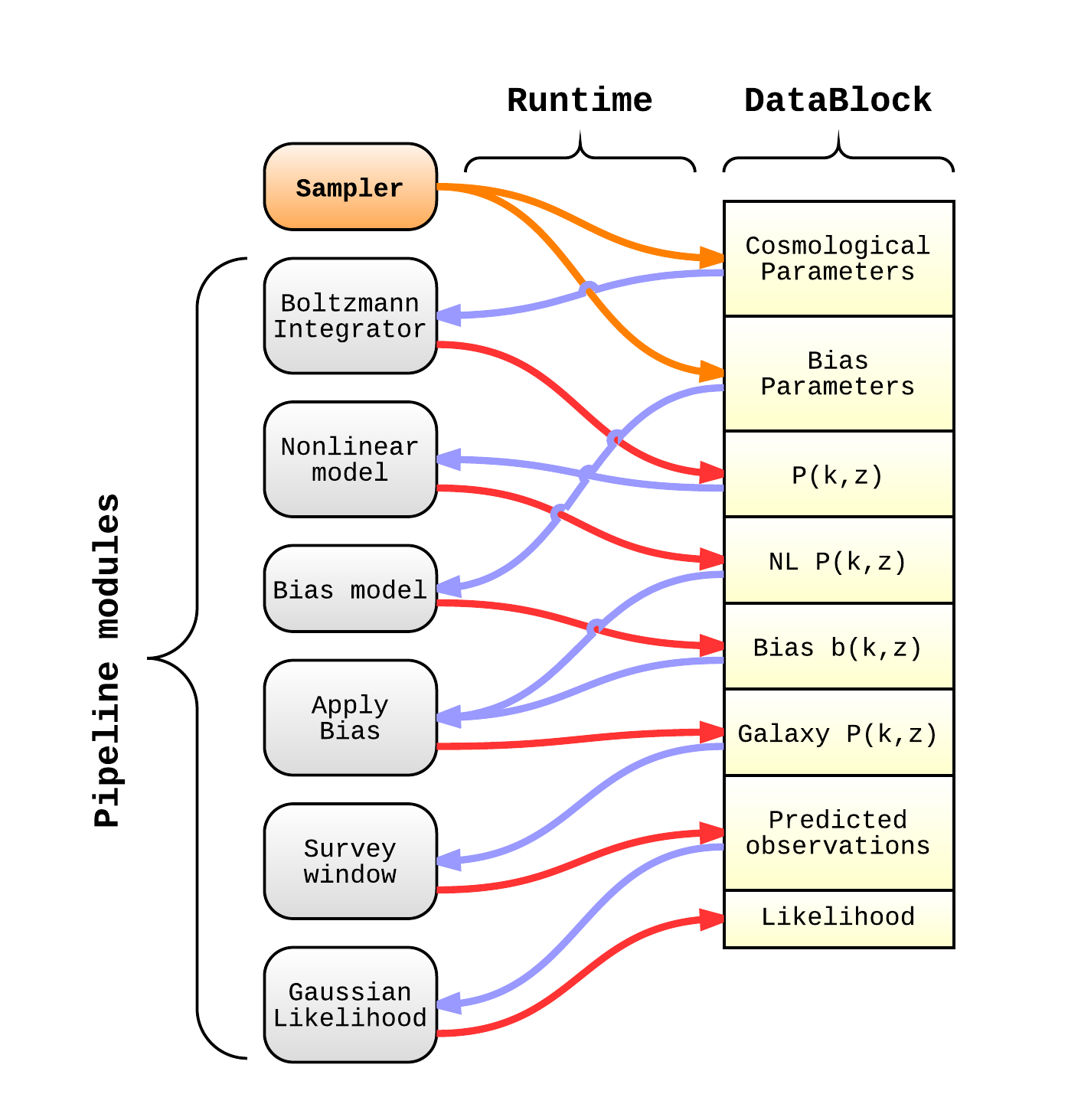}
   \caption{A schematic of the modular calculations for a galaxy power spectrum likelihood.}
	\label{galaxy power flow chart}
\end{figure}

This process is illustrated in Figure \ref{galaxy power flow chart}.

\subsection{Benefits}

There are many benefits to splitting a likelihood calculation into separate modules as shown in Figure \ref{galaxy power flow chart}:

\emph{Replacement}.  For many problems, including the first three modules shown in Figure \ref{galaxy power flow chart}, 
analysts have a choice of
 models with different parameterizations, each of which 
can be used to describe the particular physical process at each step.  
Making it easy to run different models without re-writing and recompiling code each time is easy in a modular architecture. In \cosmosis a simple change to a configuration file (or even a command line option) suffices to switch between models.

\emph{Verifiability}.  It is far easier to test individual parts of 
a long pipeline than to regression-test the whole calculation.  
In \cosmosis modules have their limitations and assumptions clearly specified, allowing analysts to create consistent pipelines that can be easily regenerated at a later date.

\emph{Debugging}.  Incorrect inputs and outputs or lack of clarity about the way pieces of code are supposed to connect 
accounts for a large fraction of software bugs.  With the \cosmosis architecture the inputs to a module are absolutely explicit and the connection between modules is clear.

\emph{Consistency}. A treatment of shared physics and systematics that is consistent across probes is essential in order to obtain accurate constraints on cosmological parameters. Writing modules that read in the values they need from the shared \cosmosis datablock rather than assuming them makes this problem explicit.

\emph{Languages}. The \cosmosis plug-in approach to adding modules 
makes it easy to switch between languages for different parts of the code.  Complicated but fast portions can be written in python so they are easier to understand, and computationally intensive portions can be in a compiled language.

\emph{Legacy}.  A wide body of disorganized but powerful code already exists in cosmology.  Wrapping parts of it as \cosmosis 
modules allows the user to include it without having to structure her own code around it.

\emph{Publishing}.  Modifications to a monolithic parameter estimation code such as {\sc cosmomc}, for example, are very unlikely to be compatible with each other or combine easily.  
For example, if one group creates a new data set with nuisance parameters
 while another makes a change to implement a theory of modified gravity, then combining those two alterations consistently is straightforward in the \cosmosis structure.

\emph{Samplers}.  Splitting likelihood calculations into 
 modules means we can create our entire \cosmosis pipeline as a more easily reusable object.  With a pipeline decoupled somewhat from the sampler, 
switching between samplers -- and therefore studying which is optimal for a particular problem --  becomes a far easier proposition.
This is an important consideration as some samplers are ineffective at fully exploring multimodal distributions or parameter degeneracies.

\subsection{Costs}

A modular structure is not free; it imposes certain costs during both the design and execution of calculations.

\emph{Overheads}. There is an overhead of code that must be written and run to connect each module to the system.  For simpler modules this can be short, but for more complex modules with multiple options it can become more difficult.  Being another layer of separation between parts of the pipeline it is also another place bugs can enter.

\emph{Interpolation}. In a monolithic CPE architecture, functions in other parts of the code can be called freely; in this modular structure 
data must be explicitly saved by one module to be useable by another.  This can mean that data is not sampled at the points that a module is expecting, and therefore require interpolation.  This can be a source of inaccuracy.  One mitigation is to define sample spacing in initialization files and check explicitly that the required accuracy is achieved, but this does place some burden on the user to perform validation tests.

\emph{Speed}. The connections between modules can be (and are, in \cosmosis) fast enough that they do not slow down cosmological likelihoods significantly. But short-cuts and efficiencies available in a tightly-coupled code may not be available in a modular context.

\emph{Consistency}. Although a modular approach can help with consistency compared to a gradually accumulated codebase it is more vunerable
to misuse compared to a rigid 
monolithic code that is  designed from the start with consistency in mind.  
This can be particularly true in complex cases such as those where errors are cosmology-dependent.
A key feature of \cosmosis is that any pipeline output contains the runtime options 
and assumptions for each module used. This makes all the parameter values and cosmological model choices explicit and allows pipelines to be regenerated easily at a later
date for verifiability or comparsion with collaborators.  
This feature limits the losses in moving away from a monolithic code and removes any ambiguity in the settings and assumptions used for a particular pipeline.

\emph{Temptation}. As it becomes easier to specify and design a pipeline the temptation to over-complicate and build large and complex pipeline grows.  The more parameters and steps a process has the more prone to error it is, and the more difficult the associated sampling problem becomes - larger spaces require more samples, longer burn-in, and make it harder to diagnose convergence.  Having powerful pipeline tools must not become an excuse to avoid thinking about how to simplify a likelihood as much as possible.

\emph{Legacy}.  Most existing code is not written with modularity in mind.  Much of it needs to be modified to fit into the \cosmosis framework\footnote{See the \cosmosis wiki for notes on importing legacy code. \url{https://bitbucket.org/joezuntz/cosmosis/wiki/modules}}. 

%% file: 3-architecture.tex
\section{CosmoSIS structure}
\label{architecture section}
In this section we provide an overview of the structure of \cosmosis and modules that link to it, and discuss the various samplers available in it in Section \ref{sampler architecture section}.  More architectural details are available in Appendiex \ref{architecture details}.

\subsection{Overview}

In \cosmosis a parameter estimation problem is represented by various components:
\begin{description}
	\item[pipeline] a sequence of calculations that computes a joint likelihood from a series of parameters.

	\item[modules] the individual ``pipes'' in the pipeline, each of which performs a separate step in the calculation.  Some do physics calculations, others interpolation, and at the end some generate likelihoods.  Many modules are shipped with \cosmosis as part of a standard library and users can easily write and include more.

	\item[datablock] the object passed down the pipeline.  For a given set of parameters all module inputs are read from the datablock and all module outputs are written to it.

	\item[sampler] (generically) anything that generates sets of cosmological and other parameters to analyze.  It puts the initial values for each parameter
 into the datablock.

	\item[runtime] the code layer that connects the above components together, coordinates execution, and provides an output system that saves relevant results and configuration.
\end{description}

The core \cosmosis datablock is written in C++ and the runtime, samplers, and user interface are written in python.  The latter was a clear choice: parsing user input and handling complex pipeline configuration and diverse other features is a field in which python excels.  The choice of writing the core in C++ was driven first by speed requirements - we never want the runtime to be dominated by read/write - and second by the flexibility that modern C++ offers - it is easy to extend the datablock to include new data types.  We also wanted a core that could be easily called all off Fortran, C, and Python, and this configuration offered an easy way to do that.

Modules can be written in C, C++, Fortran, or Python\footnote{Generally the interface to modules could be easily extended to any language that can call and be called from C.}.  

There are a number of technologies designed to connect C/C++ to python that we could have used to load and run modules, such as cython, boost-python, and swig, but we opted for a much simpler solution, the built-in \emph{ctypes} modules, a very low-level interface into shared library functions.  This was done for simplicity, clarity, and speed: ctypes is a very thin layer of abstraction, and so has minimal overhead and functions are called directly as defined.  It places a little extra burden on the cosmosis developers when writing wrapper code, since undefined behavior can occur if mistakes are made, but the reduction in compile and thinking time is large.

In ctypes shared libraries are opened by name, and functions in them are extracted by name and manually assigned argument and return types.  Once these assignments are made type checking is performed by python.  This means the end user does not need to know anything about functions exposed by ctypes in order to use them.

\subsection{Modules \& Pipelines}
\label{module architecture pipeline}
The modularity that we advocate above is embodied in the splitting of the \cosmosis likelihood function into a sequence of separate modules, each responsible for only a part of the calculation.  

A module has two parts.  The \emph{main part} of the module code performs one or more calculations that go from physical input to output. The \emph{interface} connects this to \cosmosis by loading inputs from the datablock (see below) and saving outputs back to it.  The interface is implemented either as a shared library or a python module.

Some modules exist to generate quantities for later modules to use - we refer to these as \emph{physics modules}.  Others use these values to produce data likelihoods - these are \emph{likelihood modules}.  Some can do both, and there is no structural difference between them.  A sequence of modules connected together is referred to as a \emph{pipeline}, and objects in the \cosmosis runtime manage the creation and running of pipelines and modules.

\subsubsection{Examples}

{\sc Camb} \cite{2000ApJ...538..473L} has been packaged as a \cosmosis physics module.  It loads cosmological parameters (and optionally a w(z) vector) from the datablock, and saves cosmic distance measurements and various linear power spectra.  The Planck likelihood code \cite{2013arXiv1303.5075P} has been packaged as a likelihood module - it reads CMB spectra from the datablock, and saves likelihoods.  A very simple \cosmosis pipeline could just combine these two modules.  We could substitute another Boltzmann code for {\sc camb}, such as {\sc class} \cite{2011arXiv1104.2932L}, with no changes at all to the Planck module, and compare the two just by changing a single line in a configuration file, with no recompiling.

\subsection{DataBlocks}
\label{datablocks}
We enforce modularity in \cosmosis by requiring that all information to be used by later modules is stored in a single place, which we call a \emph{DataBlock}.  Storing all the cosmology information in one places  makes it easier to serialize blocks.  It also makes debugging easier because all the inputs that a given module receive are explicitly clear.

The datablock is a \cosmosis object that stores scalar, vector, and n-dimensional integer, double, or complex data, as well as strings.  It is the object that is passed through the pipeline and contains all the physical cosmological information that is needed by the pipeline, such as cosmic distances, power spectra, correlation functions, or finally likelihoods. 

DataBlocks explicitly \emph{cannot} store arbitrary structured information; wanting to do so would suggest that modularity is broken, since genuinely physical information is typically fairly unstructured.  If complicated data structures are passed among code it would imply that code should be a single module.  A good guideline is that data stored in blocks should have physical meaning independently of the particular code that generates or uses it.

More details may be found in Appendix \ref{architecture details}.

\subsection{Samplers}
\label{sampler architecture section}

We think of a ``sampler'' in very abstract terms, and do not limit ourselves to a Markov chain Monte Carlo (MCMC).  A sampler is anything that produces one or more sets of input parameters, in any way, runs the pipeline on them, and then does something with the results.  Some samplers then iterate this process.  MCMC and maximum-likelihood samplers, for example, use previous parameter values to choose more samples, unlike a grid sampler, which decides them all in advance.

The most trivial possible sampler is implemented as the \texttt{test} sampler in \cosmosis: it generates a single set of input parameters and runs the pipeline on that one set, saving the results.

For samplers that can work in parallel, like grid sampling or {\sc emcee}, we provide a parallel ``Pool'' sampler architecture implemented with process-level parallelism using Message-Passing Interface (MPI).  Thread parallelism at the sampler level is not possible becase many key cosmology codes (like {\sc camb}) are not thread-safe.  Thread parallelism within modules is supported; for example using OpenMP.

The following samplers are available in \cosmosis. The details of how to call each sampler in a pipeline are given in section \ref{choosing_a_sampler}.
\begin{enumerate}
        \item The \texttt{test} sampler evaluates the \cosmosis pipeline at a single point in parameter space and is useful 
		for ensuring that the pipeline has been properly configured. The \texttt{test} sampler is particularly useful for generating predictions for theoretical models, outside the context of parameter estimation. 

        \item The \texttt{grid} sampler  is used to sample the CosmoSIS parameters in a regularly spaced set of points, or grid. 
		This is an efficient way to explore the likelihood functions and gather basic statistics, particularly when only 
		a few parameters are varied. When the number of parameters is large, the number of sampled points in each dimension 
		must necessarily be kept small. This can be mitigated somewhat if the grid is restricted to parameter ranges of interest.

        \item The \texttt{maxlike} sampler is a wrapper around the \texttt{SciPy minimize} optimization routine, which is by default an implementation of the Neader-Mead downhill simplex algorithm.

        \item The \texttt{metropolis} sampler implements a straightforward Metropolis-Hastings algorithm with a proposal similar to the one in {\sc cosmomc}, using a multivariate Gaussian.  Multiple chains can be run with MPI.

        \item The \texttt{emcee} \cite{2013PASP..125..306F} sampler\footnote{\url{http://dan.iel.fm/emcee/current/}} (Daniel Foreman-Mackey, David W. Hogg, Dustin Lang, Jonathan Goodman)  is a python implementation 
		of an affine invariant MCMC ensemble sampler \cite{Goodman:2010we}. The \texttt{emcee} sampler simultaneously 
		evolves an ensemble of ``walkers'' where the proposal distribution of one walker is updated based on the position of all other walkers in 
		a complementary ensemble. The number of walkers specified in the \cosmosis ini file must be even to allow a parallel stretch move where 
		the ensemble is split into two sets (see \cite{2013PASP..125..306F}). The output will be (walkers $\times$ samples) number of steps for each parameter.

		\item The \texttt{multinest} \cite{2009MNRAS.398.1601F} sampler\footnote{\url{http://ccpforge.cse.rl.ac.uk/gf/project/multinest/}}, a multi-modal nested sampler that integrates the likelihood throughout the prior range of the space using a collection of live points and a sophisticated proposal to sample in an ellipsoid containing them.  It produces the Bayesian evidence in addition to samples from the posterior.
\end{enumerate}

		A discussion of the comparative advantanges of nested, emcee, and metropolis sampling can be found in \cite{2014MNRAS.437.3918A}.

\subsection{User Interface}

The primary user interface in \cosmosis is configuration files in the ``ini'' format, extended slightly beyond the standard to allow the inclusion of other files.  The ini file is converted into a DataBlock object to initialize modules. For convenience all ini file parameters can be overridden at the command line.

%% file: 4-demos.tex
\section{Examples}
\label{examples section}

\cosmosis ships with a selection of demos that illustrate its features.  In this section we briefly overview them.

\subsection{Example One: basic cosmology functions}

\begin{figure}[h!]
	\centering
       \includegraphics[width=0.5\textwidth]{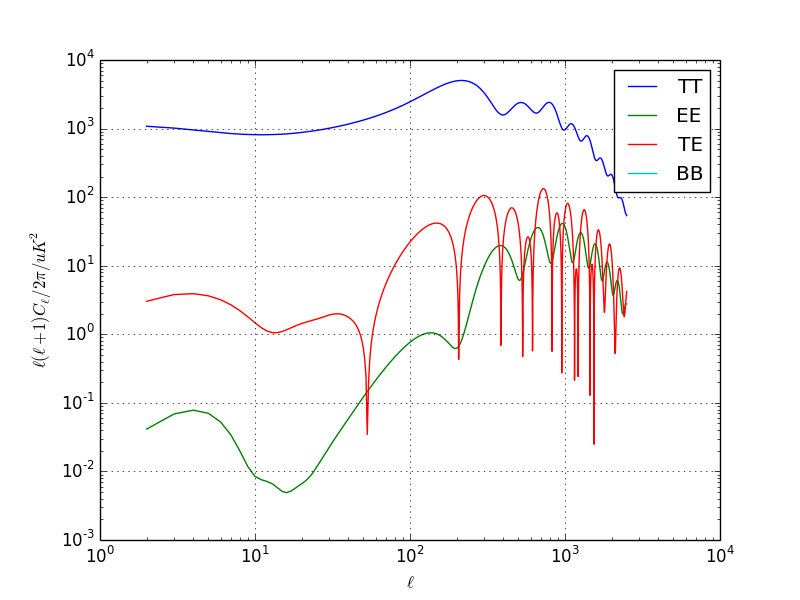}
   \caption{\cosmosis demo one output plot, showing CMB spectra output from {\sc camb}. CMB spectra and a host of other cosmology theory values are saved from the {\sc camb} \cosmosis module for later pipeline modules to use.}
	\label{demo1}
\end{figure}

The first \cosmosis demo is a simple illustration of a very basic pipeline, which produces no likelihoods and just saves some cosmological results.  The code can be run with the command \texttt{cosmosis demos/demo1.txt} and analyzed  with \texttt{postprocess demos/demo1.txt} to produce plots like Figure \ref{demo1}.

The pipeline run has just two modules - {\sc camb} and {\sc halofit}, and the results, which are saved into a newly created directory, illustrate the outputs that we extract from the two of them.

\subsection{Example Two: Planck \& BICEP2 likelihoods}

In demo two we modify demo one by adding real likelihoods to the end - for Planck and BICEP2~\cite{Ade:2014xna}.  Our pipeline is now: {\sc camb}-{\sc planck}-{\sc bicep2}, and the code can be run with \texttt{cosmosis demos/demo2.txt}.  The Planck data files are required for this demo to work.

This time our single-sample test sampler reports some output likelihood values for the pipeline.

\subsection{Example Three: BICEP2 likelihood slice}

In demo three we use our first non-trivial sampler: we take a line sample through the BICEP2 likelihood in the primordial tensor to scalar ratio $r$.  All we must do to switch to the grid sampler is change the \texttt{sampler} setting in the configuration file to \texttt{grid} and tell it how many sample points to use.

\begin{figure}[h!]
	\centering
       \includegraphics[width=0.5\textwidth]{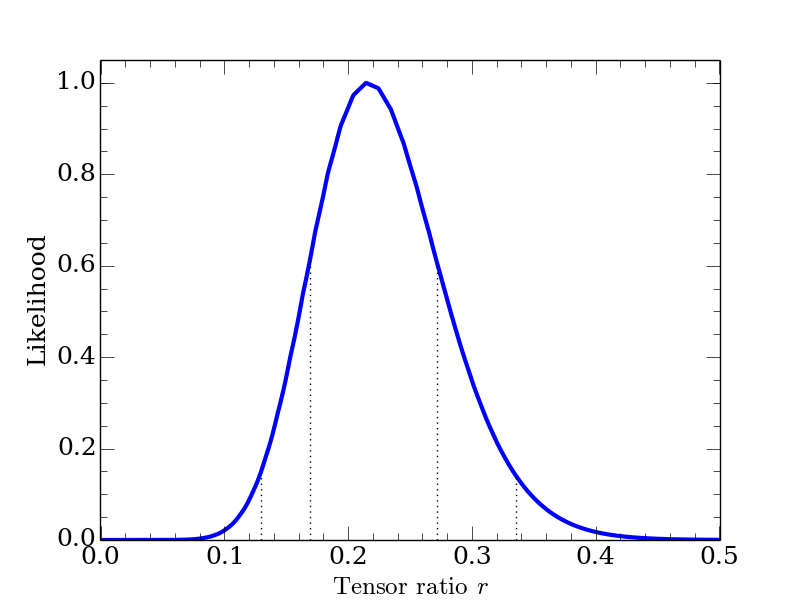}
   \caption{\cosmosis demo 3 output plot, showing the constraints on the primordial tensor fraction $r$ from BICEP2 B-mode data. The y-axis is the normalized likelihood and the vertical lines show 68\% and 95\% contours.}
	\label{demo3}
\end{figure}

Run this example with \texttt{cosmosis demos/demo3.txt} and  the results in Figure \ref{demo3} are produced with \texttt{postprocess demos/demo3.txt}, along with constraints on the $r$ parameter.

\subsection{Example Four: Maximum-likelihood Planck}

The fourth \cosmosis demo uses a numerical optimizer to find the maximum likelihood parameter set for a set of Planck likelihoods. Run it with \texttt{cosmosis demos/demo4.txt}.

The sampler uses the Nelder-Mead method \cite{Nelder:1965in} to find the peak (though various other methods can be chosen in the ini file).

The best-fitting parameters are reported at the end, and since we often use max-like samplers to find a starting point for Monte-Carlo samplers, the \cosmosis max-like sampler also outputs an ini file you can use for this purpose.  In fact demo five below starts using an ini file generated like this.

\subsection{Example Five: mcmc'ing JLA supernovae}

Demo number five brings us to geniune MCMC sampling, using the {\sc emcee} sampler.  In this pipeline we configure {\sc camb} to run only background quantities like $D_A(z)$, and then use a JLA likelihood module \cite{2014AAS...22342704B} to sample with.  We include supernova light-curve nuisance parameters.

\begin{figure}[h!]
	\centering
       \includegraphics[width=0.4\textwidth]{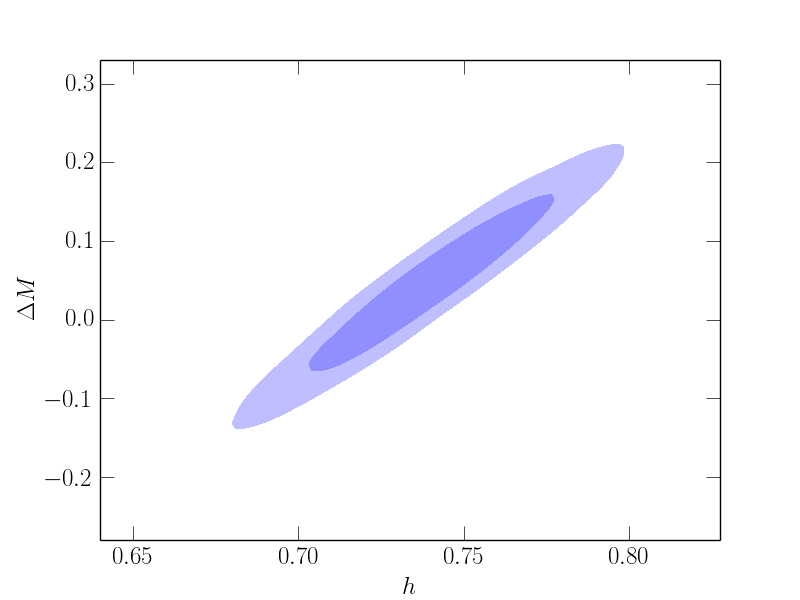}
   \caption{An example \cosmosis constraint on the JLA supernova data set -  all the 1D and 2D constraint plots are generated by \cosmosis; this example (\cosmosis demo 5) shows constraints from the JLA SDSS supernova sample on the Hubble parameter $h$ and the supernova magnitude parameter $\Delta M$ made using the \texttt{emcee} sampler.}
	\label{emcee plot}
\end{figure}

The post-process plotting code with \cosmosis generates plots from MCMCs like the one in Figure \ref{emcee plot} using kernel density estimation to smooth the samples, with a correction to ensure the right number of samples are under the 68\% and 95\% contours (see appendix \ref{kde section}).

\subsection{Example Six: CFHTLenS; a longer pipeline}

CFHTLenS is an example of a more complex likelihood pipeline of the type that will be the norm in the coming decade.  This pipeline is discussed in depth in Appendix \ref{example module}; briefly, it has six different modules: {\sc camb} - {\sc halofit} - {\sc number-density} - {\sc shear-spectra} - {\sc shear-correlations} - {\sc likelihood}.

\begin{figure*}[htpb]
	\centering
       \includegraphics[width=0.8\textwidth]{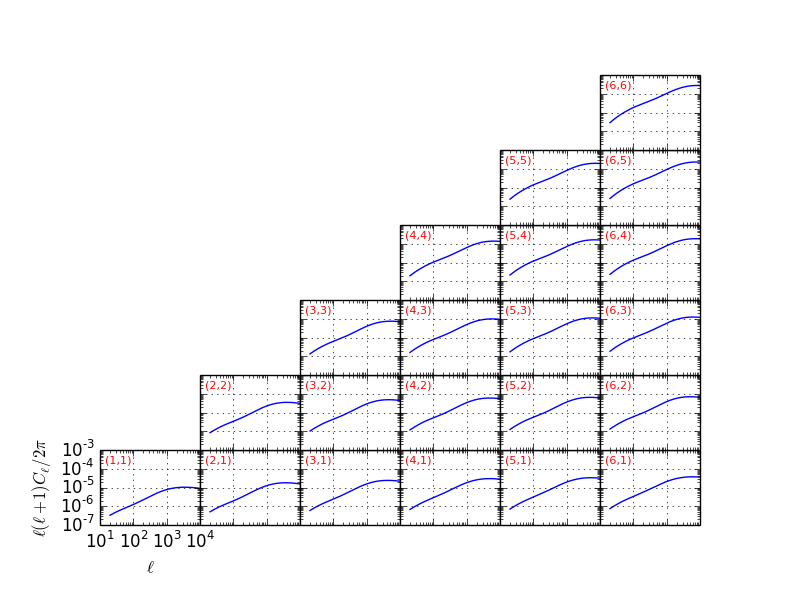}
   \caption{The \cosmosis test sampler produces and saves all the cosmological outputs for a set of parameters, and they can immediately be plotted with the postprocess program.  This example from \cosmosis demo six shows the cosmic shear spectra generated for the CFHTLenS redshift bins.}
	\label{cfhtlens plot}
\end{figure*}

Since this is quite a slow process we just run the test sampler for this demo, and produce (among other results) the plot in Figure \ref{cfhtlens plot}.  For sampling, the parallel capabilities of the {\sc emcee} and {\sc multinest} samplers are invaluable for reasonable run times.

\subsection{Example Seven: 2D grid sampling BOSS DR9}

The grid sampler from example three can grid in arbitrary dimensions\footnote{though above about 4 dimensions this becomes unfeasible, since the number of samples $ = (n_\mathrm{grid})^{n_\mathrm{dim}}$}, and in parallel if required.  In that example we used just a single dimension for a slice sample; in this example we run a 2D grid over BOSS \cite{2013MNRAS.433.3559C} constraints on the growth rate and $\sigma_8$, 
wrapping 
a \cosmosis standard library module that can calculate the linear growth rate as a function of redshift and for dynamical dark energy, $w(z)$, cosmologies.

\begin{figure}[h!]
	\centering
       \includegraphics[width=0.5\textwidth]{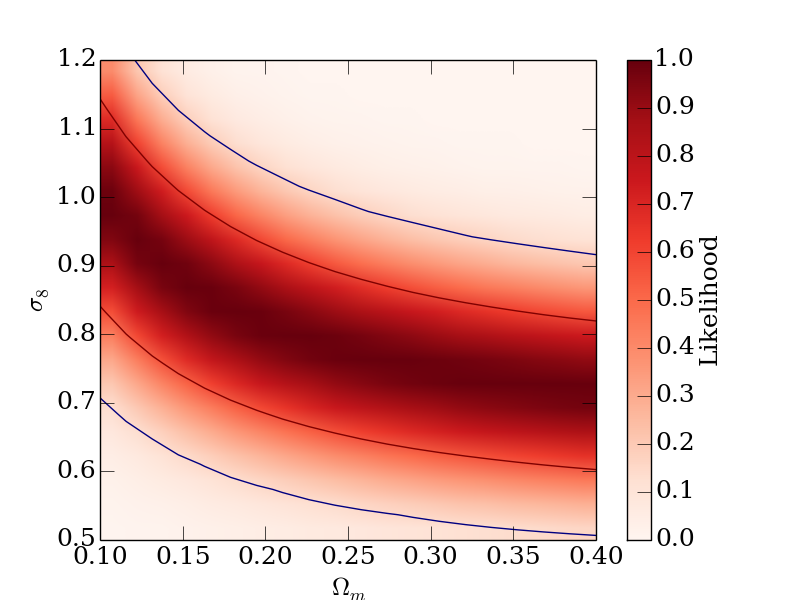}
   \caption{For smaller parameter spaces a grid sampler like the one shown here may be more suitable than an MCMC.  Grid constraints can also be immediately plotted by the \cosmosis postprocessor.  This example (from demo seven) shows contraints on $\Omega_m$ and $\sigma_8$ from BOSS measurements of $f\sigma_8$.}
	\label{grid plot}
\end{figure}

The \cosmosis post-process program reads the same ini file used to run the sampling, and thus knows automatically  that our output files are grids and that grid plots rather than MCMC constraint plots should be made.  The 2D plot for this example is shown in Figure \ref{grid plot}.  If we had sampled over other parameters they would be automatically marginalized over.

%% file: 5-sharing.tex
\section{Sharing CosmoSIS modules}
\label{sharing section}

\cosmosis comes with a collection of generally useful modules for common cosmological inference problems. We refer to this as the \cosmosis standard library (CSL), and it includes the Boltzmann code {\sc camb}; likelihoods like {\sc Planck} and {\sc CFHTLenS}; some common mass functions
from the Komatsu Cosmology Routine Library\footnote{\url{http://www.mpa-garching.mpg.de/~komatsu/crl/}} 
adapted into modules;
bias parameterizations; source count number densities; and various other calculations.

Collaborations, projects, and individuals can easily create their own libraries of \cosmosis modules to be used alongside or in place of the CSL.  These might be used to perform calculations specific to a particular theory or experiment, or for a particular paper.  They might augment the behaviour of CSL pipelines, for example by implementing a new systematic error effect, or replace standard behaviour, such as using a new improved mass function instead of a standard one. CSL is a sub-directory of the \cosmosis main directory, as are any other libraries used by collaborations and individuals.

\begin{figure}[h!]
	\centering
       \includegraphics[width=0.4\textwidth]{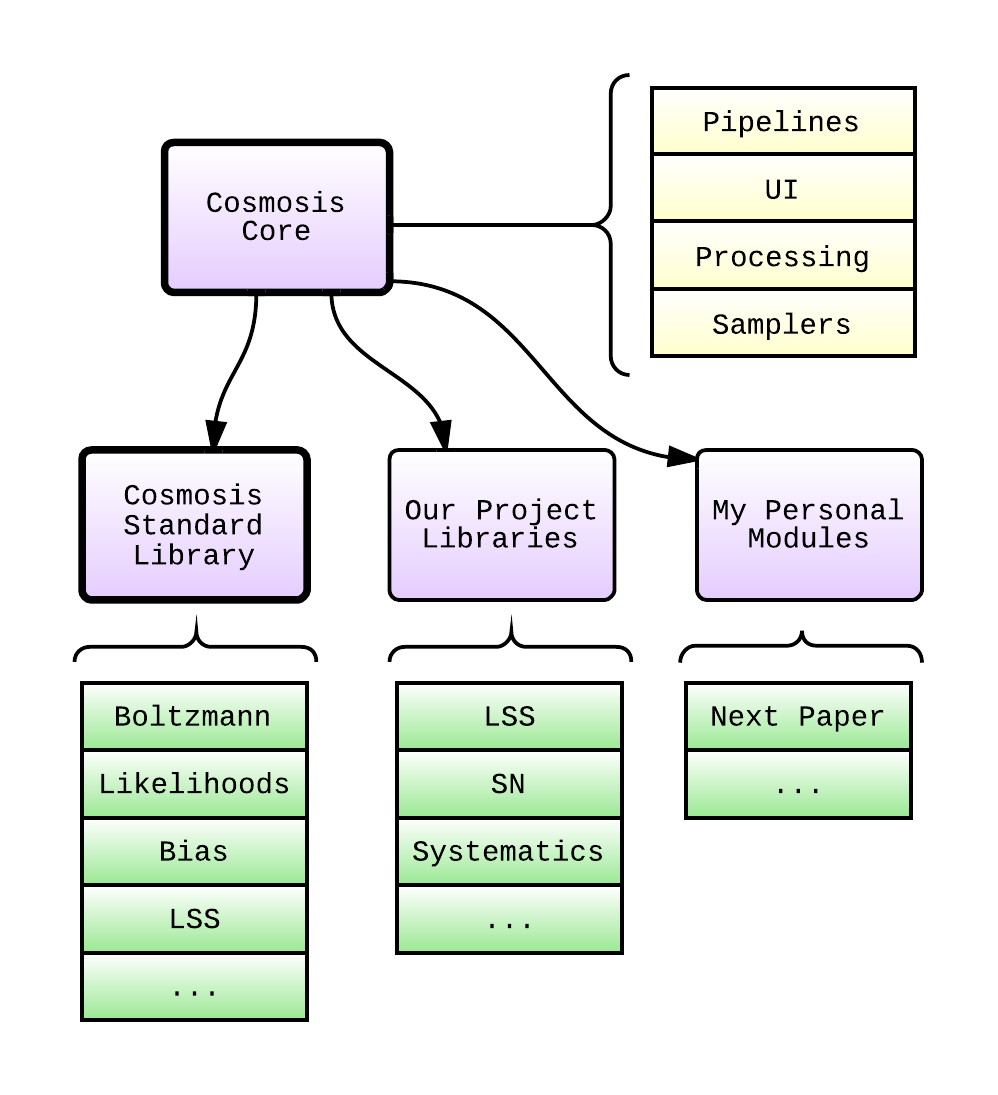}
   \caption{The structure of \cosmosis. Purple boxes are repositories, in bold if they are part of the \cosmosis package; the others are per-user.  Green boxes contain (collections of) modules.}
	\label{reposistory structure}
\end{figure}

There is exactly one sensible way to organize collections of modules: in version-controlled repositories. The repositories used in \cosmosis are described in this section and depicted in Fig.~\ref{reposistory structure}.

\subsection{Module repositories}

The CSL is stored in a publically accessible version control repository\footnote{\url{https://bitbucket.org/joezuntz/cosmosis-standard-library}}.  Such repositories store the code and a record of all the changes and additions made to it; a directory tree on disc corresponds to a remotely stored repository, and can be kept in sync with it.  

Repositories are a convenient way of storing, managing, and sharing code, and have features for reviewing and accepting third-party contributions.  A typical work pattern is to keep a repository private initially until a paper is released, and then make the module public alongside it.

Repositories can be written by and made accessible to individual collaborators, wider teams, or the community at large.

\subsubsection{Creating a new module repository}

\cosmosis comes with a script to automate the process of creating new repositories for (groups of) modules that you want to manage or share.  Run the script using \texttt{cosmosis/tools/create-repository.py}\texttt{ --help} for information on using it.

\subsection{Contributing to the standard library}

The CSL is not immutable and we very strongly welcome contributions of modules of any sort so that the cosmology community can make use of code from different groups and experiments more easily.

We will gladly distribute any module with the CSL as long as:
\begin{itemize}
\item the code authors will give permission for us to distribute and if necessary modify it.
\item any data included with it can be released publically.
\item it will be of general use to the community.
\item it meets accuracy, quality, and documentation standards.
\end{itemize}

Since most cosmologists are not trained programmers we do not enforce any specific coding standard or technology such as unit testing, but we strongly encourage contributors to write tests along with their code to ensure its functionality, and provide a mechanism within CosmoSIS to run all those tests on all modules.

Modules can be documented using the human- and machine-readable YAML format; a short YAML file included with each module describes its name, authorship, purpose, assumptions, inputs, and outputs.

\section{Discussion}
\label{discussion section}

The core claim we make in this paper is that a modular approach is useful and perhaps vital if cosmological parameter estimation is to remain accessible across the cosmology community.  We have presented \cosmosis, a code that embodies a modular architecture. It is a freely available, flexible and extendable tool for the use of observers, analysts, and theorists alike.

While the \cosmosis standard library contains a range of pre-existing and new modules for parameter estimation problems, its true value is its extensibility.  We strongly encourage and welcome contributions of existing or new code wrapped as a module\footnote{See the wiki for guidance: \url{https://bitbucket.org/joezuntz/cosmosis/wiki/modules}} and are happy to assist in any way\footnote{You can either email us or open an issue on the \cosmosis repository: \url{https://bitbucket.org/joezuntz/cosmosis/issues/new}}.

To take your next steps with \cosmosis, you can download and install the code and try the examples.  After this there are further instructions in Appendices \ref{user guide} and \ref{dev guide} on going further as a \cosmosis user and developer.

\section*{Acknowledgements}

We are grateful for help with beta testing the code to Youngsoo Park, Michael Schneider, Niall Maccrann, Matt Becker, Mustapha Ishak, Tim Eifler, Katarina Markovic, Marcelle Soares-Santos, Eric Huff, Tomasz Kacprzak.
We also thank members of the Dark Energy Survey Theory and Combined Probes Science Working Group including Ana Salvador, Marisa March, Elisabeth Krause and Flavia Sobreira.
We are grateful to the attendees of the Chicago CosmoSIS May 2014 workshop and the LSST DESC CosmoSIS Philadelphia June 2014 workshop for useful feedback.

JZ and SB acknowledge support from the European Research Council in the form of a Starting Grant with number 240672.

SD is supported by the U.S. Department of Energy, including grant DE-FG02-95ER40896. DOE HEP Computing supported development of CosmoSIS. EJ acknowledges the support of a grant from the Simons Foundation, award
number 184549. This work was supported in part by the Kavli Institute for
Cosmological Physics at the University of Chicago through grants NSF
PHY-0114422 and NSF PHY-0551142 and an endowment from the Kavli Foundation
and its founder Fred Kavli. We are grateful for the support of the
University of Chicago Research Computing Center for assistance with the
calculations carried out in this work.

This work was supported in part by National Science Foundation Grant No. PHYS-1066293 and the hospitality of the Aspen Center for Physics.

%% file: a1-user-guide.tex
\section{\cosmosis user's guide}
\label{user guide}
To run \cosmosis you:
\begin{enumerate}
	\item Choose a sequence of modules to form the pipeline.
	\item Create a parameter file describing that pipeline.
	\item Create a values file describing the numerical inputs for the parameters or their sampling ranges.
	\item Check that your pipeline can run using \cosmosis with the \texttt{test} sampler.
	\item Choose and configure a sampler, such as \texttt{grid}, \texttt{maxlike}, \texttt{multinest}, or \texttt{emcee}.
	\item Run \cosmosis with that sampler
	\item Run the \texttt{postprocess} command on the output to generate constraint plots and statistics.
\end{enumerate}

In this section we describe each of these steps in more detail.

\subsection{Choosing the pipeline}

Choosing a pipeline means deciding what cosmological analysis you wish to perform, and then selecting (or writing) a sequence of modules that realize that analysis.  This means asking the questions: what data do I want to use for my constraints?  What theory predictions do I need to compare to that data?  What calculations and parameters do I need to get that theory calculation?  And what modules do these calculations and likelihood comparisons?  

If you are using only one likelihood these questions are usually quite easy, but if there is more than one some more thought is required.  For example, if you are using only weak lensing data then you can sample directly over the $\sigma_8$ parameter, whereas if you wish to combine with CMB data you need to start with $A_s$ and derive $\sigma_8$ from it.

Typically, the pipeline will produce a likelihood, but one use of \cosmosis is to generate the theory predictions for observables for a discrete set of parameters. Working within the framework of \cosmosis  \, enables the user to exploit tools such as {\sc camb} to generate predictions.

Every module has a set of inputs and outputs, and for a pipeline to be valid every input that a module requires must be provided, either by an earlier module in the pipeline, or by the initial settings of the sampler.  Two modules must also not try to provide the same output, unless one is explicity over-writing the other, since this would imply two different and probably inconsistent methods for calculating the same thing.

Often you can write a prototypical pipeline without including various systematic errors or other steps, and then add these as new modules in the middle of the pipeline as your analysis develops.  For example, in weak lensing we must include the effect of error in our shape measurement by scaling the predicted spectra.  For an initial analysis, though, this can be left out.  A module performing the scaling can be inserted later.

\subsection{Defining the pipeline}

Once you have done the hard part and decided on a pipeline then you tell \cosmosis what you have chosen.  A \cosmosis pipeline is described in the main parameter configuration file.  Some example demos are included with \cosmosis, and modifying one of those is a good place to start.  

A section in the ini file tells \cosmosis what modules make up your pipeline, where to find values to use as inputs to it, and what likelihoods to expect at the end.  Here is an example from demo six, which analyzes tomographic CFHTLenS data \cite{2013MNRAS.432.2433H}:
\begin{lstlisting}
[pipeline]
modules = camb halofit  load_nz  shear_shear  2pt cfhtlens
values = demos/values6.ini
likelihoods = cfhtlens
\end{lstlisting}

The \texttt{modules} parameter gives an ordered list of the modules to be run.  The \texttt{values} parameter points to a file discussed in the next section.  And the \texttt{likelihoods} tells \cosmosis what likelihoods the pipeline should produce - because \texttt{cfhtlens} is listed in this case a module in the pipeline is expected to produce a scalar double value in the \texttt{likelihood} section called \texttt{cfhtlens-like} (actually the log-likelihood).  Other values in the likelihood section created by the pipeline will not automatically be included in the likelihood value for the acceptance criterion - this can be useful for importance sampling, for example.

Each entry in the \texttt{modules} list refers to another section in the same ini file, which tells \cosmosis where to find the module and how to configure it.  For example, here is the section in demo six for the first module in the pipeline, {\sc camb}:
\begin{lstlisting}
[camb]
file = cosmosis-standard-library/boltzmann/camb/camb.so
mode=all
lmax=2500
feedback=0
\end{lstlisting}

The \texttt{file} parameter is mandatory for all modules, and describes the shared library or python module interface code.  
The other parameters are specific to {\sc camb}, and are passed to it when the module is initialized.  For example, the \texttt{lmax} parameter defines the maximum $\ell$ value to which the CMB should be calculated\footnote{The parameters used by modules in the standard library of \cosmosis are described at \url{https://bitbucket.org/joezuntz/cosmosis/wiki/default\_modules}}.

Parameter files can be ``nested'': the ini file that you run \cosmosis on can use the syntax \texttt{\%include other\_params.ini} to mean that all the parameters defined in \texttt{other\_params.ini} should also be used.  This is particularly useful for running a number of similar chains with minor differences.

\subsection{Defining parameters and ranges}

In the last section we defined the pipeline and its expected outputs; in this section we define the inputs. An entry in the \texttt{[pipeline]} section of the main ini file described above was \texttt{values = demos/values6.ini}.  This file specifies the parameter values and ranges that will be sampled.  For example,
in  \cosmosis demo four, which runs a maximum-likelihood sampler on Planck data \cite{2013arXiv1303.5075P}, this file starts with:
\begin{lstlisting}
[cosmological_parameters]
omega_m = 0.2     0.3     0.4
h0      = 0.6     0.7     0.8 
omega_b = 0.02    0.04    0.06
A_s     = 2.0e-9  2.1e-9  2.3e-9
n_s     = 0.92    0.96    1.0
tau     = 0.08
omega_k = 0.0
w       = -1.0
wa      = 0.0

[planck]
A_ps_100  = 152
A_ps_143  = 63.3
A_ps_217  = 117.0
A_cib_143 = 5.0
; ...
\end{lstlisting}

The values file is divided into sections, in this case two of them, \texttt{cosmological\_parameters} and \texttt{planck}, reflecting different types of information that are stored in the datablock.  Any module can access parameters from any section. There is no pre-defined list or number of inputs; if more are required by some modules they can be freely added.

Some of the values in the file are given a single value, such as the Planck parameters and $\Omega_k$.  That indicates that for this analysis the sampler should not vary these parameters, but leave them as fixed values.  Others, like $\Omega_m$, have a lower limit, starting value, and upper limit specified.  These specifies the range of permitted values for the parameter and specifies an implicit flat prior.

Different samplers use the starting and limit values differently.  
The \texttt{test} sampler ignores the limits and just uses the 
starting value to generate a single sample. 
MCMC samplers reject proposed samples outside the range and initalize the chains at the starting value.  And the \texttt{grid} sampler uses them to specify the range of points that should be gridded.

\subsection{Test the pipeline}

As noted in Section \ref{sampler architecture section}, the simplest sampler is one that provides just a single sample.  After building a pipeline the next step is to test it with this trivial sampler.  In the \texttt{[runtime]} section of the parameter file we set the sampler to \texttt{[test]}, and then in the \texttt{[test]} section we configure it:
\begin{lstlisting}
[runtime]
sampler = test

[test]
save_dir=demo_output_1
fatal_errors=T
\end{lstlisting}

In this example we have asked the sampler to save all the data generated by all the modules in a directory structure.  This is an excellent way to check whether a pipeline is working - all the important data in the pipeline can be compared to what is expected.

Most pipelines, like all codes, will not work the first time they are run!  The test sampler also includes options to track down causes of errors, and to time code.  For convenience we also supply a simple (and easily extensible) program to plot many of the standard cosmological observables that are saved by the pipeline, to aid debugging.  An example of one such plot from \cosmosis demo six, which generates CFHTLenS likelihoods \cite{2013MNRAS.432.2433H}, is shown in Figure \ref{cfhtlens plot}.

\subsection{Choosing a sampler}
\label{choosing_a_sampler}

Different samplers produce results that are useful in different regimes.

The \texttt{grid} sampler has a number of advantages - it is straightforward to post-process, and there is no question of convergence.  It is not however, feasible to use it in more than 4 dimensions for most problems, since the number of samples grows too large.  For visualizing 1D or 2D slices in likelihood, however, we recommend it.  This can also be useful at the start of an analysis - keeping all parameters but one or two fixed to gain an intuition for the problem.

For most standard problems we recommend starting with the \texttt{maxlike} sampler to find the peak of the probability distribution, and from there\footnote{The \cosmosis max-like sampler has an option to output a values file starting from the best-fit point it finds.} running the \texttt{emcee} sampler.

For all samplers the command line usage is identical: \texttt{cosmosis [ini]} where \texttt{ [ini]} is the user specified ini file. For samplers which can be run 
in parallel (\texttt{grid} and \texttt{emcee}) the command line usage is \texttt{mpirun cosmosis --mpi [ini]}.  For technical reasons the \texttt{--mpi} flag should be ommitted when running \texttt{multinest}.
When using each of the samplers the \cosmosis ini file should contain the \texttt{[pipeline]}, \texttt{[output]} and \texttt{[module]} interface sections together with the following
sampler specific options.

When using the \texttt{ test } sampler the \cosmosis ini file should contain  the following

\begin{lstlisting}
[runtime]
sampler = test

[test]
fatal-errors = [boolean T/F]
save_dir = [output directory]
\end{lstlisting}
After execution, \texttt{output directory} will contain any data products generated during pipeline execution. If \texttt{fatal-errors} 
is set, any exceptions will cause the sampler to exit immediately.
The pipeline is evaluated at the start values for each parameter defined in values.ini.

When using the \texttt{grid} sampler the \cosmosis ini file should contain  the following

\begin{lstlisting}
[runtime]
sampler = grid

[grid]
nsample_dimension = [integer]
\end{lstlisting}
where \texttt{nsample\_dimension}  is the number of points sampled in each parameter dimension.

When using the \texttt{maxlike} sampler the \cosmosis ini file should contain  the following
\begin{lstlisting}
[runtime]
sampler = maxlike

[maxlike]
tolerance = 1e-3
maxiter = 1000   
output_ini = [output ini file]
\end{lstlisting}
The \texttt{tolerance} sets the fractional convergence criterion for each parameter; \texttt{maxiter} is the maximum number of steps to take before giving up.
If \texttt{output\_ini} is set this provides an output ini file with the best fit as the central value.
In particular the output\_ini option is useful to provide to other samplers that benefit from starting positions near the global maximum.

When using the
\texttt{metropolis} sampler the \cosmosis ini file should contain the following
\begin{lstlisting}
[runtime]
sampler = metropolis

[metropolis]
covmat = covmat.txt
samples = 100000 
Rconverge = 0.01
nsteps = 100

\end{lstlisting}
\texttt{samples} is the maximum number of samples which will be generated. The run can stop earlier than this if multiple chains are run and the \texttt{Rconverge} is set - this is a limit on the Gelman Rubin statistic.  The proposal is along eigenvectors of the covariance matrix, rotated to avoid backtracking. Of the covariance matrix is not specified in the \texttt{covmat} argument a default one will be generated based on the ranges specified in the values file.

When using the \texttt{ emcee} sampler the \cosmosis ini file should contain  the following
\begin{lstlisting}
[runtime]
sampler = emcee

[emcee]
walkers = 200
samples = 100
start_points = start.txt
nsteps = 50
\end{lstlisting}
The number of \texttt{walkers} must be at least $2*{\rm nparam} + 1$, but in general more than that usually works better (default  = 2); 
\texttt{samples} is the number of steps which each walker takes (default = 1000) and sample points are output every \texttt{nsteps} (default = 100).
The starting points for each walker in the chain may be specified in the ini file using \texttt{start\_points = start.txt} where start\_file.txt 
contains (number of walkers, number of params) values. If this start file is not given then all walkers are initialized with uniform random numbers from the range of values in values.ini. 
For practical Monte Carlo the accuracy of the estimator is given by the asymptotic behaviour of its variance in the limit of long chains. Forman-Mackney et al. advocate the following:
Examining the acceptance fraction which should lie in the range 0.2-0.5. Increasing the number of walkers can improve poor acceptance fractions.
Estimating the autocorrelation time which is a direct measure of the number of evaluations of the posterior probability distribution function 
required to produce independent samples of the target density.

\subsection{Running \cosmosis}

Regardless of the sampler or other parameter choices, \cosmosis is run through a single executable invoked on the configuration file.  MPI parallelism is enabled at the command line flag (in combination with any mpirun command required), and any parameter in the configuration files can be over-written using another flag (this feature is mainly useful for debugging).

\subsection{Processing outputs}

A post-processing program for sampler output, simply called \texttt{postprocess} uses the output of chain and grid samplers to generate 1D and 2D marginalized constraint plots and numbers.  You call it on the same ini file that was used to generate the chain in the first place, so that any type of chain (grid, mcmc, or any others that we add) are analyzed with the same executable.

An example output of the \texttt{postprocess} command on the \texttt{emcee} sampler is shown in Figure \ref{emcee plot}, and from the grid sampler in Figure \ref{grid plot}.

%% file: a2-dev-guide.tex
\section{Developers's Guide}
\label{dev guide}
Most users of parameter estimation codes go on to modify and write their own code. Making this easy is the whole point of \cosmosis.

There are several ways you can develop within \cosmosis:
\begin{itemize}
\item Modify modules in an existing pipeline, for example to study new physics.
\item Add modules that do different physical calculations than what is currently available.
\item Add a new likelihood to the end of a pipeline to combine new datasets.
\item Insert a module into the middle of a pipeline, for example to model a new systematic error.
\item Start from scratch creating a new pipeline using the \cosmosis structure and samplers.
\item Add a new sampler and test it on existing problems.
\end{itemize}

\subsection{Creating new modules}
\label{creating new modules}
If no existing module does the calculation you need, or if you wish to wrap an external piece of existing code, then you can create a new module for it.  Each new module should live in its own directory.

Unless a module is exceedingly simple it is best for new modules to be in two parts - the part that does the calculation, and the part that interfaces with \cosmosis.  In the case of wrapping existing code the former usually exists already and only the latter needs to be created.

To write a module it suffices to write the three functions described in Section \ref{module architecture pipeline}.  This involves thinking carefully what the inputs and outputs will be to this module, and deciding which of the inputs will definitely be fixed throughout a run (for example, a path to a data file, a redshift at which an observation has been made, or a choice of which model to use), and which are those which may at some point be varied throughout a chain (such as cosmological parameters).

Appendix \ref{example module} shows an example of a simple module.

\subsubsection{Module form}

Modules implemented in python are connected to \cosmosis using a single python file that implements the functions described below.  Modules implemented in a compiled language (C, C++, or Fortran) must be compiled into a shared library, which can be loaded dynamically from python.  This simply involves compiling all files and linking them together with the \texttt{-shared} flag.  Examples can be found in the \cosmosis standard library.

\subsubsection{setup}

The setup function is called once, at the start of a run when a pipeline is being created.  It is a chance to read options from the \cosmosis parameter files, allocate memory and other resources, and load any data from file.  If a distributed parallel sampler is used this may mean being called by each processor separately.

The setup function is passed a datablock object which contains all the information from the parameter file.  In particular, options set in the section of the ini file corresponding to this particular module have a specifically named section;  modules need only look at this section.  The API for accessing the data from the configuration file is described in Appendix \ref{API}.

The setup function can return any arbitrary configuration information, which is then passed back to the exectue function below.  This mechanism means that the same module can be run twice with different options (for example, the same likelihood module can be used with two different data sets).

\subsubsection{execute}

The execute function is the main workhorse of a module - it is called each time the sampler runs the pipeline on a new set of parameters. The execute function takes two parameters, one containing the parameters from the sampler and any data from modules earlier in the pipeline, and one containing the configuration information from the setup function.

A typical module reads some scalar parameters or vector data from the block, and then performs some calculations with it depending on the choices made in the ini file.  It then saves new scalar or vector data back to the block. Appendix \ref{API} describes the API for loading and saving values.

\subsubsection{cleanup}

A cleanup function is run when the pipeline is finished, and should free any resources loaded by the setup function.  In many cases this can be completely empty.  This function is passed the configuration information from the setup function.

\subsubsection{New likelihoods}

Any module can, as well as doing any other calculations, save values into the \texttt{likelihoods} section.  This section has a special meaning - the samplers will search in it for any likelihoods that they are told to find in the parameter file.  If the parameter file says, for example:
\begin{lstlisting}
[pipeline]
likelihoods = hst planck
\end{lstlisting}
then the sampler will look for \texttt{hst\_like} and \texttt{planck\_like} in the \texttt{likelihoods} section.

\subsection{Modifying existing modules}

When you want to test a new theory it is usually easiest to take an existing module and modify it to implement the new theory.  For example, the \cosmosis module to calculate the growth function could be changed to implement a modified gravity scenario.

Modifying an existing module to extend it by add new calculations, rather than modifying existing ones, is usually a sub-optimal choice, since new calculations can be better integrated with other modules if they are in a separate module.  Consider writing a new module instead.

You would take these steps to modify an existing module:
\begin{itemize}
\item Copy the existing module to a new location.
\item Version control your new module.
\item Modify the main module science code.
\item Modify the interface code if any new inputs or outputs are required.
\end{itemize}

\subsection{Inserting modules}

If you want to make a modification to a quantity from a physical effect that a pipeline does not currently consider, then you can insert a new module in the middle of the pipeline to implement it.

For example, it is known that baryons have a feedback effect on the matter power spectrum which is important for observations probing non-linear cosmic scales.  
One might insert a module after e.g., the Halofit module, to modify the non-linear matter power to account for this effect, so that subsequent modules would use the modified power.

Inserted modules can be created as described in Section \ref{creating new modules}, but there is one additional consideration.  The \cosmosis DataBlock (see Section \ref{datablocks}) makes a distinction between saving new values and replacing existing ones.  Modules that modify existing data need to use the \texttt{replace} operations described in Appendix \ref{API} instead of the \texttt{put} ones.

\subsection{Adding samplers}

New samplers can be easily added to \cosmosis if they can be called from python; this includes any sampler usable as a library (e.g. with a few simple functions that can be called to run the sampler with an arbitrary likelihood function) in C, C++, or Fortran, as well as python samplers.

Interfaces to samplers are implemented by subclassing a Sampler base class, in the \texttt{cosmosis/samplers} subdirectory.  The subclasses must implement three methods:
\begin{itemize}
\item \texttt{config}, which should read options from the ini file, and perform any required set up.  The superclass instance has an instance of the ini file used to create the run and the instantiated pipeline itself.
\item \texttt{execute}, which should perform a single chunk of sampling, and saving the result - superclass methods can be used for the latter.  \cosmosis will keep re-running the \texttt{execute} function until the sampler reports it is converged.
\item \texttt{is\_converged}, which should return True if the sampling should cease.  A simple sampler might always return True, but a more complex sampler can examine the chain so far and use real convergence diagnostics.
\end{itemize}

Parallel samplers inherit instead from the ParallelSampler superclass, which as well as Sampler's features maintains a pool of processes, each with their own instance of the pipeline, and an \texttt{is\_master} method to decide if the given process is the root one.

%% file: a3-example-code.tex
\section{Worked example}
\label{example module}
In this appendix we show a worked example of a \cosmosis pipeline, and go into detail about one of the modules in it.  

Our example will be \cosmosis demo number six, which calculates the likelihood of the CFHTLenS tomographic data set given some cosmological parameters.

\subsection{Overview}
The CFHTLenS observed data vector is a set of measurements of $\xi_+$ and $\xi_-$, correlation functions of the cosmic shear.  We have measuremnts $\xi^{ij}_{\pm}(\theta_k)$ for each pair of redshift bins\footnote{Since CFHTLenS is a photometric experiment the redshifts are approximate, so the actual redshift distribution in each bin is different from the nominal one.  We must account for this in the analysis.} $(i,j)$ and for a range of angular scales $\theta_k$.

The steps we need to take to calculate the likelihood from the cosmological parameters are therefore:
\begin{enumerate}
\item Calculate the linear matter power spectrum $P(k,z)$ across the desired redshifts
\item Calculate the non-linear power spectrum from the linear
\item Calculate the redshift distribution of the survey
\item Perform the Limber integral to get the shear angular power spectra.
\item Integrate the angular power spectra with Bessel functions to get the angular correlation function.
\item Get the likelihood of the CFHTLenS measurements given these theory correlation functions.
\end{enumerate}

\begin{figure}[h!]
	\centering
       \includegraphics[width=0.5\textwidth]{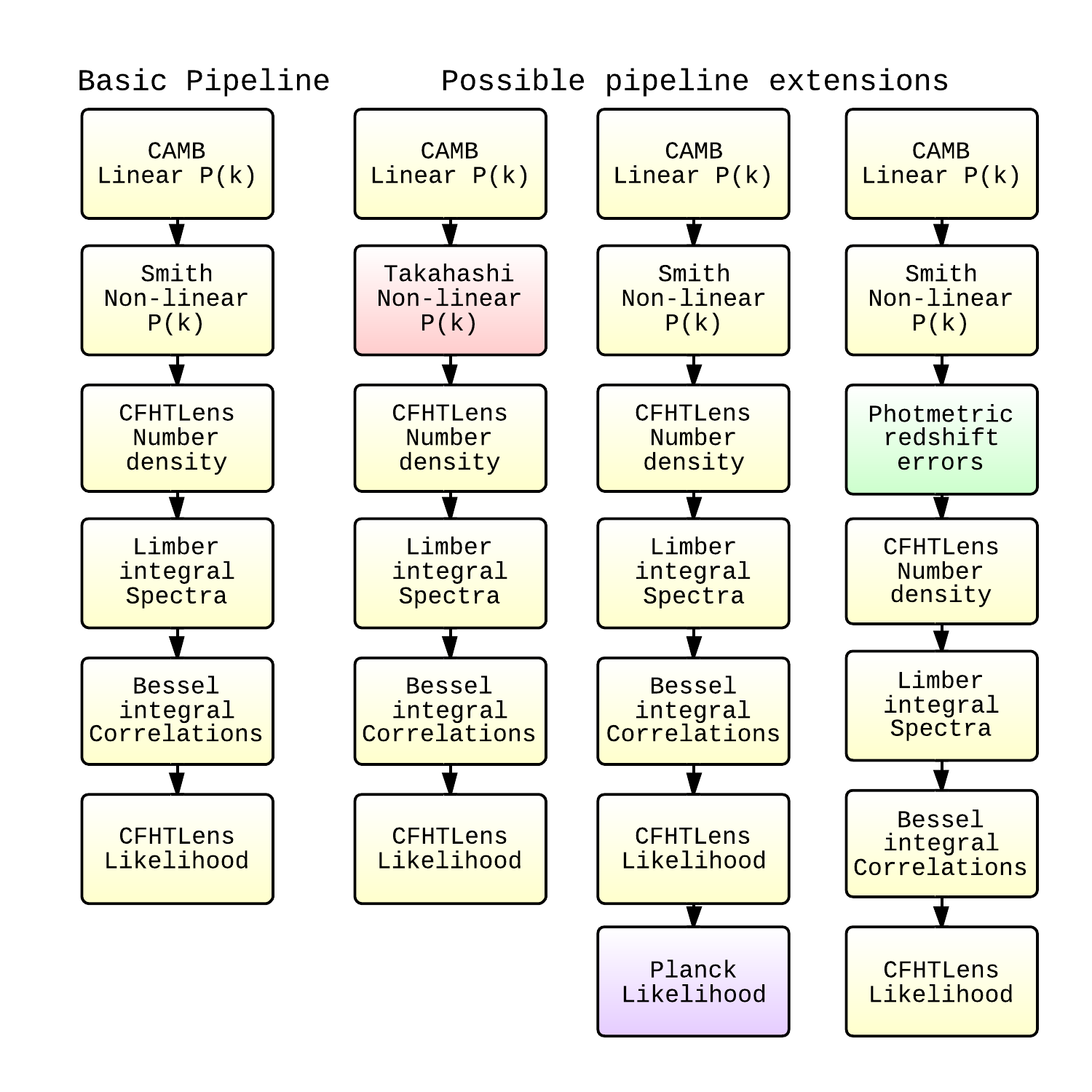}
   \caption{A schematic of the CFHTLenS pipeline. The basic pipeline is on the left, and in the three alternatives on the right we illustrate replacing (red), appending (purple) and inserting (green) modules.}
	\label{cfhtlens pipelines}
\end{figure}

This pipeline is shown in Figure \ref{cfhtlens pipelines}.

\subsection{Modifications}
Figure \ref{cfhtlens pipelines} also shows various changes we might wish to make to this pipeline; making it easy to implement these changes is a core goal of \cosmosis.

The simplest example is replacing one module with another, in this case changing the method used to calculate non-linear power. Provided each module supplies the same outputs, subsequent modules can be left unchanged by this replacement.

We can also straightforwardly attach new likelihoods to the end of the pipeline, illustrated here by adding the Planck likelihood.  This likelihood requires new nuisance parameters, which can be supplied by the sampler simply by adding them to the input values file.

Finally, we might insert a module into the middle of the pipeline, in this case to test for a systematic error.  If we decided, for example, to model errors in the photometric redshifts determined by CFHTLenS then we could modify and replace the $n(z)$ used in the spectra.

\subsection{Pipeline implementation}
Each box in Figure \ref{cfhtlens pipelines} is a single \cosmosis module.  Each does a single calculation and gets a relatively small collection of inputs from the DataBlock, and puts its outputs there once they are calculated.

The configuration file that runs this module needs to define the sampler to be used, just the test sampler in this case:

\begin{lstlisting}
[runtime]
sampler = test

[test]
save_dir=demo6_output
fatal_errors=T
\end{lstlisting}

and also the modules to be run and the likelihoods extracted:
\begin{lstlisting}
[pipeline]
modules = camb halofit  load_nz  shear_shear  2pt cfhtlens
values = demos/values6.ini
likelihoods = cfhtlens
\end{lstlisting}

The \texttt{value} parameter lists a file where the values and ranges of the parameters to be sampled are specified.  Each module to be run is described elsewhere in thre configuration file, in some cases with extra options:

\begin{lstlisting}
[camb]
file = cosmosis-standard-library/boltzmann/camb/camb.so
mode=all
lmax=2500
feedback=0

[halofit]
file = cosmosis-standard-library/boltzmann/halofit/halofit_module.so
...
\end{lstlisting}

\subsection{Module implementation}

The simplest module in this pipeline is the one that performs the integration with Bessel functions to convert angular power spectra to correlation functions.  In this section we will describe the interface that connects this module to \cosmosis in detail.

This particular module is implemented in python, but similar (if slightly more complex) considerations apply to the other supported languages.

\subsubsection{Preamble}

We will not delve here into the implementation of the main workhorse of this module; we simply import it into python.  Note that we have separated the main functionality of the code from the part that connects it to \cosmosis; this make it easy, for example, to use the same code in other programs easily

\begin{lstlisting}
import cl_to_xi
import numpy as np
from cosmosis import option_section, names as section_names
\end{lstlisting}

\subsubsection{Initial setup}

The setup function is run once when the pipeline is created.  Options from the configuration file are passed into it as the \texttt{options} object, which is a DataBlock.  The \texttt{option\_section} is a shorthand for the section that applies to this particular module (modules can in principle find out what other modules are to be run also).

The various required options (which concern the angular ranges at which to calculate correlations) are read here and the range constructed.

Anything returned by this function will be passed to the \texttt{execute} function laster.  In this case that means a dictionary, \texttt{config}, that contains the vector \texttt{theta}, over which to compute the correlation functions.

\begin{lstlisting}

def setup(options):
	config = {}
	n_theta = options[option_section, "n_theta"]
	theta_min = options[option_section, "theta_min"]
	theta_max = options[option_section, "theta_max"]
	theta_min = cl_to_xi.arcmin_to_radians(theta_min)
	theta_max = cl_to_xi.arcmin_to_radians(theta_max)
	theta = np.logspace(np.log10(theta_min), np.log10(theta_max), n_theta)
	config["theta"] = theta

	return config
\end{lstlisting}

\subsubsection{Execution}

This function is called each time the pipeline is run with new cosmological parameters.  The \texttt{block} input contains the parameter-specific information: the values provided by the sampler itself, and the calculation results done by previous modules.  The \texttt{config} input contains fixed data passed from the \texttt{setup} function (though this could in principle be modified, for example to cache results).

It reads the inputs from the pipeline section \texttt{shear\_cl}, which are the ell range \texttt{ell} provided by the modules that came before it, and the \texttt{bin\_1\_1}, etc., giving the angular power spectra.  The results are saved into \texttt{shear\_xi}.

\begin{lstlisting}

def execute(block, config):
	thetas = config["theta"]
	n_theta = len(thetas)

	section = section_names.shear_cl
	output_section = section_names.shear_xi
	ell = block[section, "ell"]
	nbin = block[section, "nbin"]
	block[output_section, "theta"] = thetas
	block.put_metadata(output_section, "theta", "unit", "radians")

	for i in xrange(1,nbin+1):
		for j in xrange(1,i+1):
			name = "bin_%d_%d"%(i,j)
			c_ell = block[section, name]
			xi_plus, xi_minus = cl_to_xi.calculate_xi(ell, c_ell, thetas)
			block[output_section, "xiplus_%d_%d"%(i,j)] = xi_plus
			block[output_section, "ximinus_%d_%d"%(i,j)] = xi_minus
	return 0
\end{lstlisting}

\subsubsection{Clean up}
In python there is rarely any clean up to be done, since memory is managed automatically.  In C or Fortran you might deallocate memory here.

\begin{lstlisting}

def cleanup(config):
    return 0
\end{lstlisting}

%% file: a4-architecture-details.tex
\section{Architectural Details}
\label{architecture details}
\subsection{DataBlocks}

DataBlocks are organized into sections, named categories of information.  For example, \texttt{cosmological\_parameters}, \texttt{cmb\_cl} and \texttt{intrinsic\_alignment\_parameters} can all be sections.  A number of common sections are pre-defined in \cosmosis, but they are simple strings and new ones can be arbitrarily created.  A datablock may be thought of as a dictionary mapping from a pair of strings (a section name and a specific name for data in that section) to a generic value.

For example, the \texttt{cosmological\_parameters} section would typically contain \texttt{omega\_m}, \texttt{h0}, and other scalar doubles, and the \texttt{cmb\_cl} section contains an integer array \texttt{ell} and double arrays \texttt{TT}, \texttt{TE}, \texttt{EE}, and so on.

Native APIs that act on datablocks exist for C, Fortran, C++, and Python to read or write the data stored in the block. The interfaces to modules (see below) call this API, as do the samplers when they create the block in the first place.

There are also introspection functions that work on datablocks so that modules can peform context-dependent calculations (for example, we might check for a galaxy bias grid $b(k,z)$ and if one is not found revert to a single scalar $b$ value).

\subsection{Samplers}

Samplers are connected to \cosmosis by sub-classing from a base class which provides access to the pipeline and to configuration file input, and to output files.  Subclasses 
implement methods to \texttt{config} (read options from the ini file and perform setup) \texttt{execute} (run a chunk of samples) and test \texttt{is\_converged} to see if the process should stop.  Adding a new sampler is straightforward and we would welcome contributions.

%% file: a5-technical.tex
\section{API}
\label{API}
The \cosmosis application programming interface (API) defines a way for a module to save and load data from a block designed to collect together all the theoretical predictions about a cosmology.  The API is fully documented on the \cosmosis wiki\footnote{\url{https://bitbucket.org/joezuntz/cosmosis/wiki}}.

The API can handle the following types of data:
\begin{itemize}
\item 4-byte integers
\item 8-byte floating-point (real) values
\item 4-byte boolean (logical) values
\item ASCII strings
\item $8+8$ byte complex numbers
\item vectors of 4-byte integers
\item vectors of 8-byte floats
\item vectors of $8+8$-byte complexes
\item n-dimensional arrays of 4-byte integers
\item n-dimensional arrays of 8-byte floats
\item n-dimensional arrays of $8+8$-byte complexes
\end{itemize}

Any value stored in a block is referenced by two string parameters, a section defining the group in which it is stored, and a name of the value.  For each type in each supported programming language there are \texttt{get}, \texttt{put} and \texttt{replace} functions.  There are also a number of additional utility functions to check whether values exist in the block, and similar tasks.

In this section we show a handful of available API calls to demonstrate their general structure.

\subsection{C}

Most  C functions return type \texttt{DATABLOCK\_STATUS}, an enum.  For each type listed above there are function to \texttt{get}, \texttt{put}, and \texttt{replace} values.  For the scalar types there are also alternative \texttt{get} functions where a default value can be supplied if the value is not found.  In the case of 1-d array there are two \texttt{get} functions, one to use preallocated memory and one to allocate new space, which the module is reponsible for disposing of.  For the n-d arrays there are also functions to query the number of dimensions and shape of the array.

The function to get an integer, for example, has this prototype:
\begin{lstlisting}
DATABLOCK_STATUS
c_datablock_get_int(c_datablock* s, const char* section, const char* name, int* val);
\end{lstlisting}

\subsection{Fortran}
The Fortran functions closely follow the C ones, and use the \texttt{iso\_c\_binding} intrinsic module to define types.  For example:
\begin{lstlisting}
function datablock_get_double(block, section, name, value) result(status)
    integer(cosmosis_status) :: status
    integer(cosmosis_block) :: block
    character(*) :: section
    character(*) :: name
    real(c_double) :: value
\end{lstlisting}

\subsection{Python}
While \texttt{get\_int} and similar values are present in the python API, the most straightforward mechanism to load and save values is the idiomatic python get and set syntax:
\begin{lstlisting}
	block["section_name", "value_name"] = value
\end{lstlisting}

All the python functions are methods on a DataBlock object.

\subsection{C++}

In C++ the \texttt{get}, \texttt{put}, and \texttt{replace} functions are all templated methods on a \texttt{DataBlock} object, so the same method is used for all data types, for example:
\begin{lstlisting}
template <class T>
DATABLOCK_STATUS put_val(std::string section, std::string name, T const& val);
\end{lstlisting}

%% file: a6-kde.tex
\section{Improving 2D KDE}
\label{kde section}
Kernel density estimation (KDE) is a method for smoothing a collection of MCMC samples to produce a better constraint plot.  It can be applied in any number of dimensions, and can be thought of as placing a smooth Gaussian (or other) kernel atop each MCMC sample and using the sum of all these Gaussians as the likelihood surface.  The main choice to be made is the covariance matrix (or just width in one dimension) of the kernel, which is typically taken as some scaling factor times the covariance matrix of the samples.

An occasional objection to KDE is that the recovered contours drawn on the smoothed distribution do not typically contain the correct fraction of samples (68\%, 95\%, etc.) that they should do if the samples and the contours accurately represented the same posterior surface.

The \cosmosis post-processing code implements uses KDE with a minor improvement when used in 2D.  The 2D likelihood surface is generated as in normal KDE.  The contours drawn on them, though, are not chosen with reference to the probability volume beneath the smoothed contours, but rather by interpolating so that the correct number of samples from the MCMC is beneath them.  That is, the KDE provides the \emph{shape} of the contours, but the sample count provides their \emph{size}.  We find that this procedure improves the fidelity of the recovered contours.

%% file: a7-consistency.tex
\section{Parameter consistency \& alternate specifications}

In cosmology, as in most parameter estimation problems, there are a number of different parameterizations one can use to specify the space.  The choice affects how easy it is to specify parameter priors, and how efficent sampling in the space can be (for most algorithms parameters with roughly Gaussian posteriors make for better sampling).

In some cases deducing the ``derived'' parameters from the specified ones requires complex calculations (for example, getting $\sigma_8$ from $A_s$) but in other cases the relations are relatively simple arithmetic.

\cosmosis includes a module with an algorithm for the latter case which allows one to specify any sufficient combination of parameters, and deduce the rest, no matter which combination they are in.  The steps of this algorithm are:

\begin{enumerate}
\item Specify a comprehensive collection of relations between parameters as strings, for example \texttt{omega\_m} \texttt{ = omega\_c+omega\_b}, \texttt{omega\_c =} \texttt{omega\_m-omega\_b}, etc.  Call the number of relations $n$.
\item Parse the left-hand side of the relations to get a set of all the parameters to be calculated.
\item Initialize a dictionary of all these parameters with the special value $\mathrm{NaN}$ (not-a-number) for each of them.
\item For any parameters which are provided by the user, initialize with the specified value.
\item Iterate at most $n$ times:
\begin{enumerate}
\item Evaluate each relation in the collection, with the current parameters using the python \texttt{eval} function with the parameter dictionary as the namespace.  There are three cases:
\begin{itemize}
\item If the result is $\mathrm{NaN}$, do nothing - this means at least one input to the relation was unspecified ($\mathrm{NaN}$).
\item If the result is not $\mathrm{NaN}$ and the current parameter value is $\mathrm{NaN}$, then we have a newly calculated parameter.  Update the dictionary with this value.
\item If the result is not $\mathrm{NaN}$ and the current parameter value is not $\mathrm{NaN}$, then we have re-calculated the parameter.  Check that this new calculation is the same as the old one.  If not, raise an error: the model is over-specified.
\end{itemize}
\item If there are no $\mathrm{NaN}$ parameter left then we have finished; save all the parameters.
\end{enumerate}
\item If we evaluate all the relations $n$ times without calculating all parameters then there are some we cannot calculate - the model is under-specified and we raise an error.
\end{enumerate}